\documentclass[aps,balancelastpage,superscriptaddress,sort&compress,longbibliography,floatfix]{revtex4-1}
\usepackage{graphicx}
\usepackage{dcolumn}
\usepackage{bm}
\usepackage{amsmath}
\usepackage{amssymb}
\usepackage{latexsym}
\usepackage{epsfig}
\usepackage{amsbsy}
\usepackage{array}
\usepackage{amssymb}
\usepackage{setspace}
\usepackage{multirow}
\usepackage{threeparttable}
\usepackage{subfig}
\usepackage{tikz}
\usepackage{textcomp}
\usepackage[export]{adjustbox}

\def\sint{\ifmmode{- \!\!\!\!\!\! \int}
    \else{\hbox{$- \!\!\!\! \int \ $}}\fi}

\makeatletter
\newcommand*{\rom}[1]{\expandafter\@slowromancap\romannumeral #1@}
\makeatother

\makeatletter
\renewcommand{\fnum@figure}{Figure \thefigure}
\makeatother

\usepackage{lineno}

\begin{document}



\title{Theory of transformation-mediated twinning}

\author{Song Lu\footnote{songlu@kth.se (S.L.)}}
\affiliation{Applied Materials Physics, Department of Materials Science and Engineering, Royal Institute of Technology, Stockholm SE-100 44, Sweden}
\author{Xun Sun\footnote{xunsun@kth.se (X.S.)}}
\affiliation{Frontier Institute of Science and Technology, State Key Laboratory for Mechanical Behavior of Materials, Xi'an Jiaotong University, Xi'an, 710049, China}
\affiliation{Applied Materials Physics, Department of Materials Science and Engineering, Royal Institute of Technology, Stockholm SE-100 44, Sweden}
\author{Xianghai An}
\affiliation{School of Aerospace, Mechanical \& Mechatronic Engineering, The University of Sydney, Sydney, NSW 2006, Australia}
\author{Wei Li}
\affiliation{Applied Materials Physics, Department of Materials Science and Engineering, Royal Institute of Technology, Stockholm SE-100 44, Sweden}
\author{Yujie Chen}
\affiliation{School of Aerospace, Mechanical \& Mechatronic Engineering, The University of Sydney, Sydney, NSW 2006, Australia}
\affiliation{School of Mechanical Engineering, University of Adelaide, Adelaide, SA 5005, Australia}
\author{Hualei Zhang \footnote{hualei@xjtu.edu.cn (H.L.Z.)}}
\affiliation{Center of Microstructure Science, Frontier Institute of Science and Technology, State Key Laboratory for Mechanical Behavior of Materials, Xi'an Jiaotong University, Xi'an, 710049, China}
\author{Levente Vitos}
\affiliation{Applied Materials Physics, Department of Materials Science and Engineering, Royal Institute of Technology, Stockholm SE-100 44, Sweden}
\affiliation{Department of Physics and Astronomy, Division of Materials Theory, Uppsala University, Box 516, SE-75121, Uppsala, Sweden}
\affiliation{Research Institute for Solid State Physics and Optics, Wigner Research Center for Physics, Budapest H-1525, P.O. Box 49, Hungary}

\begin{abstract}
High-density and nanosized deformation twins in face-centered cubic (fcc)materials can effectively improve the combination of strength and ductility. However, the microscopic dislocation mechanisms enabling a high twinnability remain elusive. Twinning usually occurs via continuous nucleation and gliding of twinning partial dislocations on consecutive close-packed atomic planes. Here we unveil a completely different twinning mechanism being active in metastable fcc materials. The transformation-mediated twinning (TMT) is featured by a preceding displacive transformation from the fcc phase to the hexagonal close-packed (hcp) one, followed by a second-step transformation from the hcp phase to the fcc twin. The thriving nucleation of the intermediate hcp phase is driven by the thermodynamic instability and the negative stacking fault energy of the metastable fcc phase. The intermediate hcp structure is characterized by the easy slips of Shockley partial dislocations on the basal planes, which leads to both fcc and fcc twin platelets during deformation, creating more twin boundaries and further enhancing the prosperity of twins. The fundamental understanding of the dislocation mechanism of deformation twinning in metastable alloys paves the road to design novel materials with outstanding mechanical properties.
\end{abstract}

\maketitle




\section{Introduction}

Plastic properties of metallic materials are mainly mediated by the nucleation and motion of dislocations. When plastic deformation is solely accommodated by dislocation slips, the strength-ductility dilemma is generally inevitable since the conventional metallurgical strengthening methods via generating internal barriers for dislocation motion cause a sacrifice in ductility \cite{Lu2009,Robert}. Introducing additional deformation mechanisms like strain-stimulated phase transformation or twinning is an effective strategy for attaining both high strength and excellent ductility as successfully demonstrated in the so-called transformation-induced plasticity (TRIP) and twinning-induced plasticity (TWIP) alloys \cite{Li227,Lu1804727,DECOOMAN2018283}. In the classical theory of plasticity for face-centered cubic (fcc, $\gamma$) metals, deformation twinning (DT) is realized by layer-by-layer nucleation and gliding of Shockley partial dislocations on consecutive  close-packed  planes; while deformation-induced martensite transformation (DIMT) from fcc to hexagonal close-packed (hcp, $\varepsilon$) structure ($\gamma\rightarrow\varepsilon$) by regular partial movements on every other close-packed planes. Since the partial dislocations for DT and DIMT are identical, in the extant theories of plasticity as well in practice \cite{Jo06052014,DECOOMAN2018283}, the above two mechanisms are considered exclusive to each other, operating in materials with different ranges of stacking fault energies (SFEs) or local SFEs due to chemical variations \cite{Niu1363,Ding8919,Chen2020}. Empirically, a critical measured SFE value ($\sim$10-20 mJ m$^{-2}$, depending on chemistry \cite{DECOOMAN2018283}) is often placed for the deformation mode transition from DIMT to DT. However, in many metallic systems \cite{WU2005681,REMY1976123,Remy1974,SHEN2012514,REMY197799,CHOI2015391,SHEN20136093,Li227,Lu1804727,Lin2018236,Niu1363,Slone38,Miao35,GUO2019176},
DIMT and DT have been simultaneously observed in the same grains, forming a unique alternate lamellar structure composed of nano-thicknessed fcc twin ($\gamma_{\rm tw}$) and hcp laths. Such nanoscaled deformation structure is crucial, because both the $\gamma/\gamma_{\rm tw}$ twin and the $\gamma/\varepsilon$ phase boundaries can not only carry significant plasticity, but also interact strongly with dislocations, catalyzing the remarkably enhanced work-hardening capacity \cite{Lu2009}.

However, the concurrence of DIMT and DT challenges the current knowledge about the underlying microscopic mechanism, which remains elusive even after about half a century since its early observation \cite{Remy1974}.  The highly unsatisfying understanding of the deformation dynamics in these metals and alloys has limited the ability to provide an accurate prediction of the composition-structure-property relationships, and restricted the capability of designing materials with superior mechanical properties. 

Here we zoom in the transition zone between DT and $\gamma\rightarrow\varepsilon$ DIMT to shed light on the atomistic processes responsible for the deformation mode change. We identify an unorthodox twinning mechanism in thermodynamically unstable (metastable) fcc metals and alloys which does not follow the classical  $\gamma\rightarrow\gamma_{\rm tw}$ layer-by-layer twinning (cTW) path. Using quantum mechanical density functional theory (DFT) calculations, we show that in such systems DT can be realized through an intermediate $\gamma\rightarrow\varepsilon$ DIMT by sequentially activated groups of partial dislocations which lead to a two-step $\gamma\rightarrow\varepsilon\rightarrow\gamma_{\rm tw}$ twinning process. This novel transformation-mediated twinning (TMT) mechanism originates from the metastability of the fcc lattice and from the basal plane shear instability of the hcp lattice. Remarkably, the TMT mechanism can provide a profound understanding for the pronounced twinning activities in metastable alloys. We pinpoint to the critical factors controlling the deformation mode transition from DIMT to DT, which enable alloy design targeting exceptional mechanical properties to maximally harvest the TWIP/TRIP effect. 

\section{Methods}

Ab initio calculations based on DFT were performed using the exact muffin–tin orbitals method \cite{vitos2007computational}. The exchange-correlation functional was described with the generalized gradient approximation \cite{perdew1996generalized}. The Kohn–Sham equations were solved within the scalar-relativistic and soft-core schemes. The chemical and magnetic disorders were taken into account using the coherent potential approximation \cite{vitos2007computational}. The free energies at room temperature included the lattice expansion and magnetic entropy terms. For metals and alloys with Curie temperature higher than room temperature,  ferromagnetic calculations were performed; whereas systems with low Curie temperatures were described in the paramagnetic state modeled by the disordered local magnetic moment approximation \cite{gyorffy1985first}. 

The $\gamma$-surfaces in both fcc and hcp lattices were calculated using the hexagonal supercells with 12 close-packed atomic layers along the $\bf c$ ($<$111$>_{\rm fcc}$ or $<$0001$>_{\rm hcp}$ ) direction. The generalized stacking fault structures were obtained by tilting $\bf c$ axis along the $<$11\={2}$>_{\rm fcc}$ or $<$\={1}010$>_{\rm hcp}$ directions, respectively, by a shear vector $\bf u$, i.e., $\bf c =\bf c +\bf u $, where $\bf u$ is from 0 to one Burgers vector of the Shockley partial dislocation. The numerical parameters were set so that the error bar in the computed intrinsic material parameters is below $\sim$2 mJ m$^{-2}$.

\section{Results}

 \begin{figure}[h]
\subfloat{{\includegraphics[trim=1cm 0cm 8cm 0cm,clip, scale=0.35]{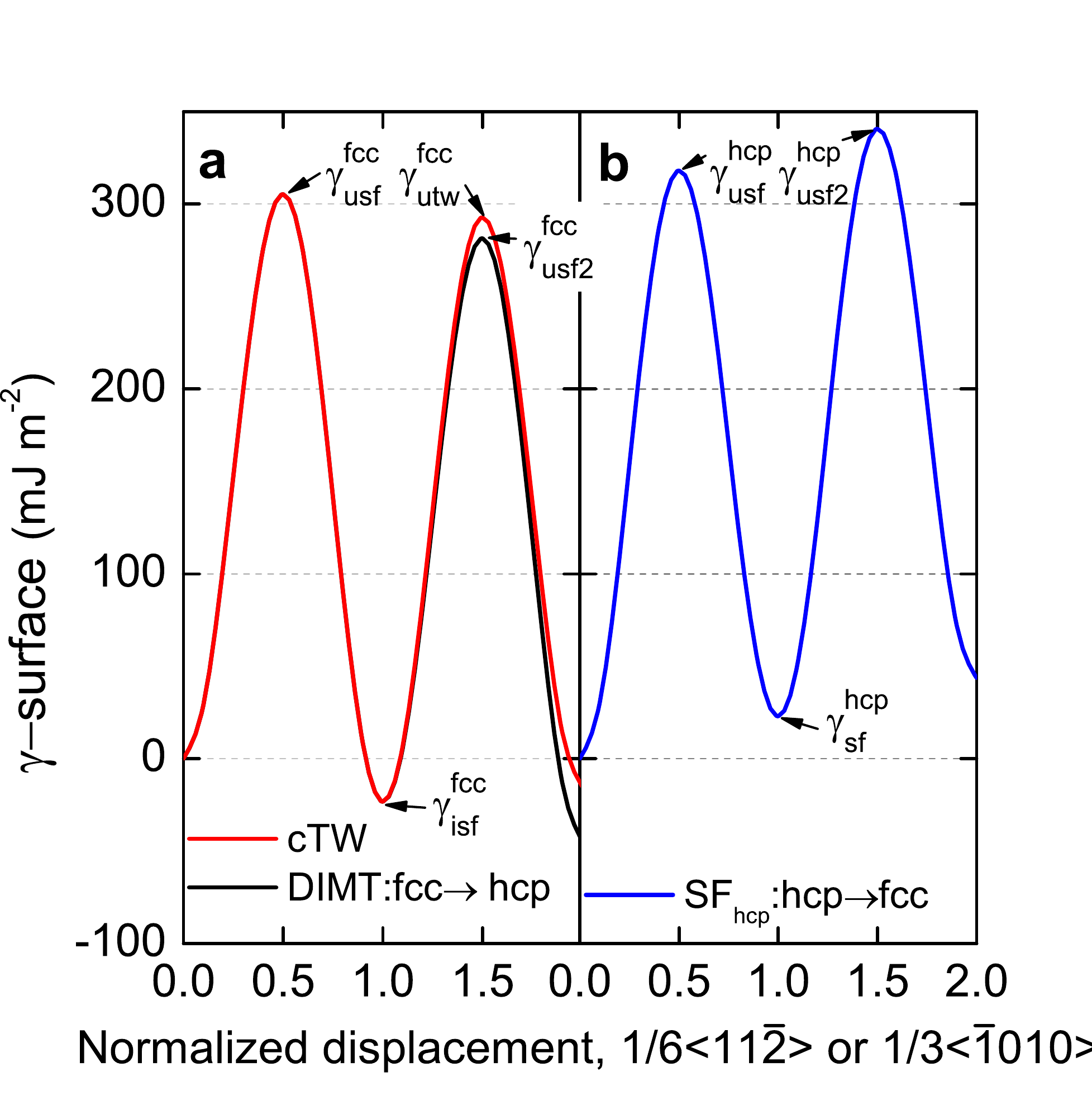}}}%
 \subfloat{{\includegraphics[trim=1cm 0cm 2cm 0cm,clip,scale=0.35]{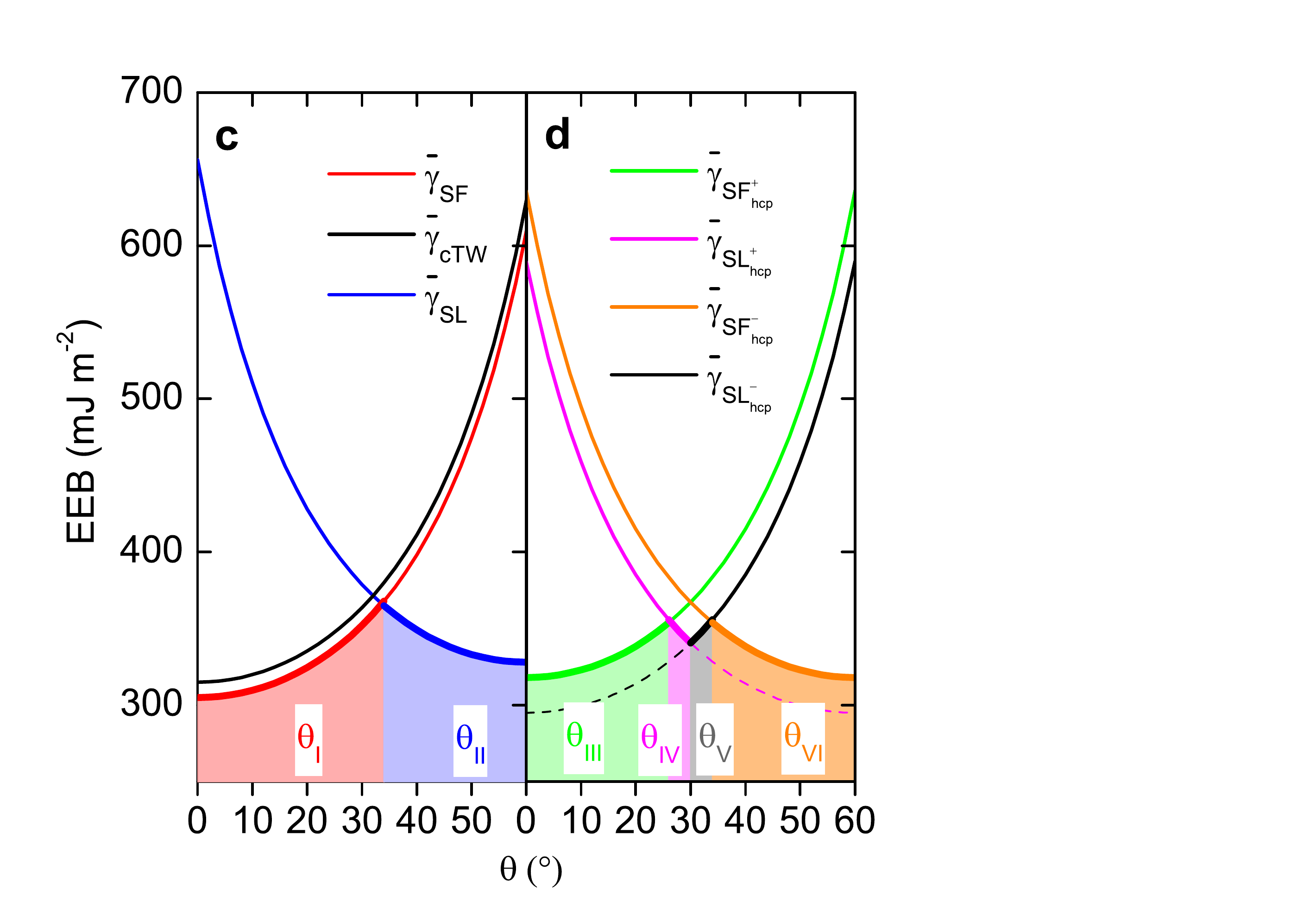}}}%
 
 \subfloat{{\includegraphics[trim=0cm 11cm 0cm 3cm,clip,width=0.8\textwidth,]{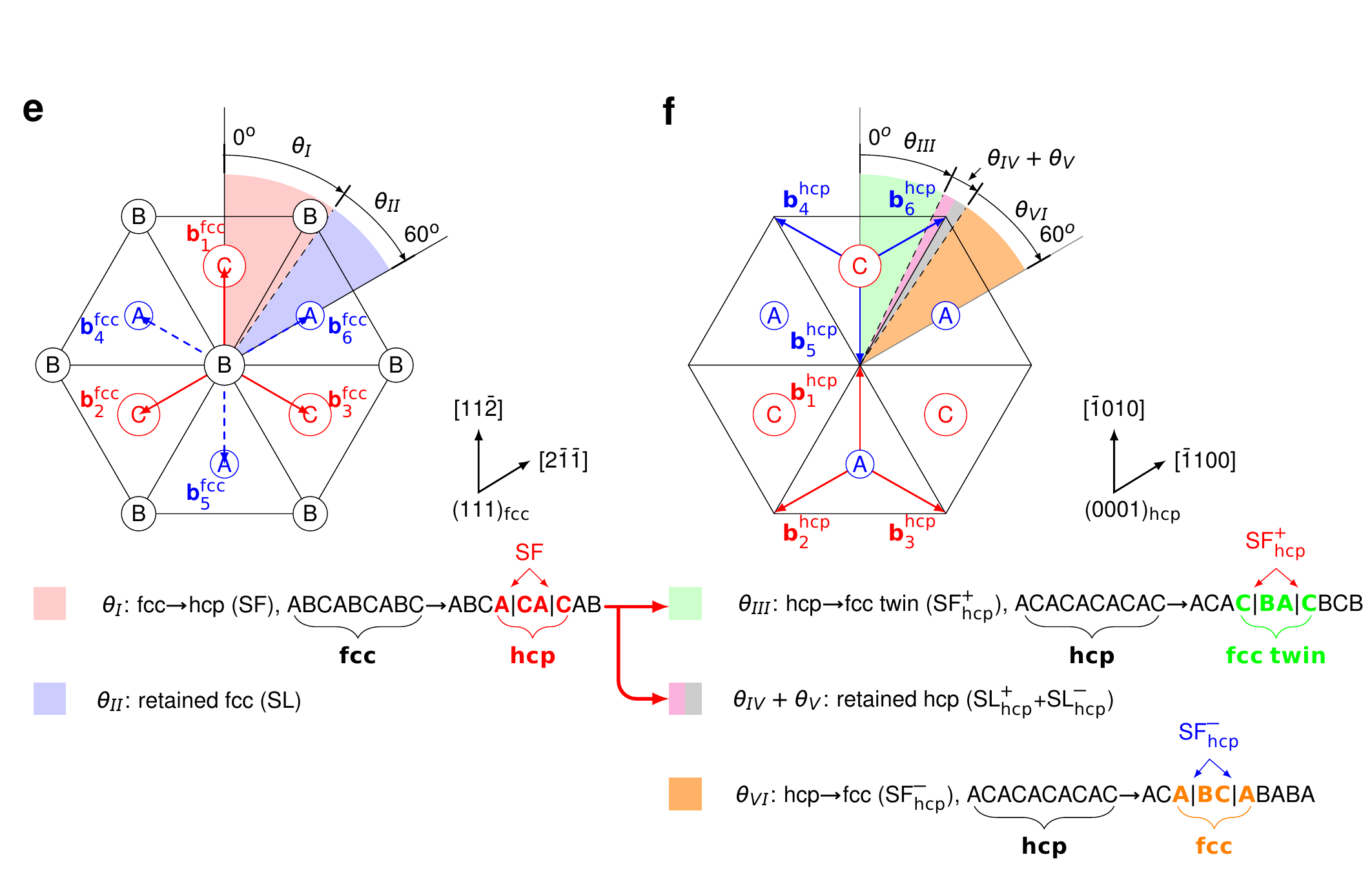}}}%
 \caption{{\bf $\gamma$-surfaces and energy barriers in fcc and hcp CrCoNi}. {\bf a} and {\bf b}, The room temperature $\gamma$-surfaces for $\gamma\rightarrow\varepsilon$ DIMT and cTW in fcc and for the basal stacking fault (SF$_{\rm hcp}$) formation in hcp lattices, respectively. {\bf c}, The orientation dependent EEBs for SF formation, cTW, and SL in fcc lattice. {\bf d}, The orientation dependent EEBs for stacking faults (SF$^+_{\rm hcp}$ and SF$^-_{\rm hcp}$) and basal dislocation slips (SL$^+_{\rm hcp}$ and SL$^-_{\rm hcp}$) on the basal planes in hcp lattice. The lowest EEBs corresponding to the favored deformation modes are highlighted by thick lines. {\bf e} and {\bf f}, The orientation dependent deformation modes are illustrated on the (111)$_{\rm fcc}$ planes in fcc and the (0001)$_{\rm hcp}$ basal planes in hcp lattices, respectively.}%
\label{figure1}%
\end{figure}

\subsection{ Classical twinning versus $\gamma\rightarrow\varepsilon$ martensitic transformation}

We take the extensively studied metastable CrCoNi medium-entropy alloy (MEA) to investigate the competition between $\gamma\rightarrow\varepsilon$ DIMT and DT, but emphasize that the discussion presented here applies to most of the metastable metals and alloys. This alloy possesses a number of intriguing mechanical properties due to its remarkable twinnability, especially at cryogenic conditions \cite{Gludovatz2016,LAPLANCHE2017292}. The generalized stacking fault energies (GSFEs, $\gamma$-surfaces) obtained for the fcc  and hcp phases of CrCoNi are shown Figs.\ref{figure1}a and \ref{figure1}b, respectively. The calculated fcc SFE is negative at room temperature ($\gamma^{\rm fcc}_{\rm isf}= -21$ mJ m$^{-2}$), which signals that the hcp structure is energetically favored. This is in line with previous theoretical results \cite{Zhang2017,Niu1363,Huang20182381}, as well as the experimental observations showing a large amount of stacking faults (SFs) already at early deformations \cite{Niu1363,Slone38,Miao35,LAPLANCHE2017292}. Furthermore, starting from an existing SF, the hcp nucleation and growth have a lower energy barrier ($\gamma_{\rm usf2}^{\rm fcc}$) to overcome than that for the classical layer-by-layer twinning nucleation (unstable twinning energy, $\gamma_{\rm utw}^{\rm fcc}$, Fig.\ref{figure1}a). Theoretically, the competition among deformation modes in a fcc material can be quantitatively measured by the effective energy barriers (EEBs) based on the $\gamma$-surface \cite{Jo06052014,Huang20182381}. The EEBs for SF formation (the same as $\gamma\rightarrow\varepsilon$ DIMT), cTW, and full dislocation slip (SL) were defined as 


\begin{linenomath*}
\begin{equation}
\overline{\gamma}_{\rm SF}(\theta)=\gamma^{\rm fcc}_{\rm usf}/\cos{\theta}, \;
\overline{\gamma}_{\rm cTW}(\theta)=(\gamma^{\rm fcc}_{\rm utw}-\gamma^{\rm fcc}_{\rm isf})/\cos{\theta},\;
\text{and } \overline{\gamma}_{\rm SL}(\theta)=(\gamma^{\rm fcc}_{\rm usf}-\gamma^{\rm fcc}_{\rm isf})/\cos{(60^{\circ}-\theta)},
\end{equation}
\end{linenomath*}
respectively, where $\gamma^{\rm fcc}_{\rm usf}$ is the unstable SFE. Here, $\theta$ measures the angle between the resolved shear stress on the \{111\}$_{\rm fcc}$ slip planes and the Burgers vector for leading/twinning partials (e.g., $a$/6[11\={2}] on (111)$_{\rm fcc}$ plane) to account for the unidirectional nature of DT/DIMT in fcc structure \cite{McCabe2014,GUTIERREZURRUTIA20103552}. $\theta$ spans 0$^{\circ}$ - 60$^{\circ}$ according to the symmetry of the \{111\}$_{\rm fcc}$ planes. The preferred deformation mode is decided by the lowest EEB \cite{Jo06052014}. 
We must emphasize here that the above EEB for twinning ($\overline{\gamma}_{\rm cTW}(\theta)$) is defined according to the classical layer-by-layer twinning route \cite{Jo06052014,Huang20182381}, therefore, $\overline{\gamma}_{\rm cTW}(\theta)>\overline{\gamma}_{\rm SF}(\theta)$ implies that energetically leading partial dislocations will avoid nucleation and gliding on nearest neighboring slipping planes. This is the case of fcc CrCoNi (Fig.\ref{figure1}c). Crystal orientation can change the preference of SF and SL through altering the resolved shear stresses on specific slip systems, i.e., different Schmid factors. From Fig.\ref{figure1}c, we see that SF and $\varepsilon$ formation occurs for small $\theta$ ($\theta_{\rom{1}}$= 0$^{\circ}$-34$^{\circ}$) and SL is the favored deformation mode at larger $\theta$ ($\theta_{\rom{2}}$= 34$^{\circ}$-60$^{\circ}$). However, the competition between cTW and  $\gamma\rightarrow\varepsilon$ DIMT is not affected by $\theta$ because they are accomplished through the same type of partial dislocations. Overall, the prediction that DIMT is preferred over twinning is apparently at odds with the indisputable observation of a large amount of deformation twins. Extant theories of twinning nucleation based on the cTW path,  when modified by considering local chemical variations \cite{Niu1363,Ding8919,Chen2020}, a substantial affine shear \cite{Huang20182381}, or local adiabatic heating \cite{Lu1804727}, may turn twinning favorable locally in multicomponent solid solutions, but all these scenarios fail to explain the observed low fraction of $\varepsilon$ martensite, and the characteristic arrangement of $\gamma/\gamma_{\rm tw}/\varepsilon$ lamellar structure \cite{Niu1363,Slone38,Miao35,LAPLANCHE2017292,GUO2019176}, nor they can be universally applied to explain similar observations in other metals with negative SFE like fcc Co ($\gamma^{\rm fcc}_{\rm isf}=-97$ mJ m$^{-2}$, Supplementary Table 1) \cite{WU2005681}.  

\subsection{Phase transformation mediated twinning}

We observe that  $\gamma\rightarrow\varepsilon$  DIMT itself is not an effective avenue for strain accommodation, considering that only partials on every other slip planes are involved in the process. Therefore, the fresh deformation-induced $\varepsilon$ lamella formed in the fcc matrix is also subject to strains during further deformation, which is indeed evidenced by the pronounced dislocation activities observed in the $\varepsilon$ martensite, particularly at large strains \cite{Li227}. Due to the strictly complied Shoji-Nishiyama (S-N) orientation relationship (OR) \cite{Miao35}, i.e., \{111\}$_{\rm fcc}$//\{0001\}$_{\rm hcp}$ and $<$11\={2}$>$$_{\rm fcc}$//$<$\={1}010$>_{\rm hcp}$, the dominant dislocation activities occur on the hcp basal planes \cite{Bu075502}. However, due to the ...ACAC... stacking, fundamental difference arises for the activation of the leading partials in the hcp lattice compared to that in the fcc one, regarding the crystal orientation dependence. In the fcc structure (stacking sequence ...ABCABC...), on all three consecutive \{111\}$_{\rm fcc}$ planes there is only one set of 1/6$<$11\={2}$>$ Burgers vectors ($ {\bf b}_{1}^{\rm fcc}$, $ {\bf b}_{2}^{\rm fcc}$ and $ {\bf b}_{3}^{\rm fcc}$, deviated by 120$^{\circ}$, Fig.\ref{figure1}e) for the leading partials which can create SFs. In contrast,  in the hcp lattice there are two sets of Burgers vectors for the leading partials gliding on the basal planes C ($ {\bf b}_{1}^{\rm hcp}$, $ {\bf b}_{2}^{\rm hcp}$ and  $ {\bf b}_{3}^{\rm hcp}$) and A ($ {\bf b}_{4}^{\rm hcp}$, $ {\bf b}_{5}^{\rm hcp}$ and $ {\bf b}_{6}^{\rm hcp}$), which deviate from each other by 60$^{\circ}$ (Fig.\ref{figure1}f). In other words, there are six equivalent glide directions on the hcp basal planes to generate SFs (denoted as SF$_{\rm hcp}$). Explicitly, in the fcc lattice, the atomic movement carried out by the leading partials is always A$\rightarrow$ B$\rightarrow$ C$\rightarrow$ A; whereas the reverse slip sequence A$\rightarrow$ C$\rightarrow$ B$\rightarrow$ A  along the trailing/anti-twinning directions is prohibited by the significantly higher slip barrier. On the other hand, in the hcp lattice, the A$\rightarrow$ B$\rightarrow$ C$\rightarrow$ A and A$\rightarrow$ C$\rightarrow$B$\rightarrow$ A  slip sequences are equivalent and can be realized on the C and A slip planes, respectively.  In accordance with the hcp stacking sequence, under a specific resolved shear stress, the leading partial dislocations with the same Burgers vector are restricted to slip on every other (0001)$_{\rm hcp}$ planes, which actually leads to the $\varepsilon \rightarrow \gamma$ phase transformation. We assign  the  A$\rightarrow$ B$\rightarrow$ C$\rightarrow$ A and A$\rightarrow$ C$\rightarrow$B$\rightarrow$ A slip sequences with the  + and - signs, respectively. On each hcp basal plane, the SF$_{\rm hcp}$ formation competes with the full dislocation slip (denoted as SL$_{\rm hcp}$) and the corresponding EEBs for the +/- slip sequences are  
 \begin{linenomath*}
\begin{equation}
\overline{\gamma}_{\rm SF^{+}_{\rm hcp}}(\theta)=\gamma^{\rm hcp}_{\rm usf}/\cos{\theta},\;
\overline{\gamma}_{\rm SL^{+}_{\rm hcp}}(\theta)=(\gamma^{\rm hcp}_{\rm usf}-\gamma^{\rm hcp}_{\rm sf})/\cos{(60^{\circ}-\theta});
\end{equation}
\end{linenomath*}
 and 
 \begin{linenomath*}
\begin{equation}
\overline{\gamma}_{\rm SF^{-}_{\rm hcp}}(\theta)=\gamma^{\rm hcp}_{\rm usf}/\cos{(60^{\circ}-\theta}),\;
\overline{\gamma}_{\rm SL^{-}_{\rm hcp}}(\theta)=(\gamma^{\rm hcp}_{\rm usf}-\gamma^{\rm hcp}_{\rm sf})/\cos{\theta}.
\end{equation}
\end{linenomath*}
Here, $\gamma^{\rm hcp}_{\rm sf}$ and $\gamma^{\rm hcp}_{\rm usf}$ are the hcp stable and unstable SFEs, respectively (Fig.\ref{figure1}b). $\theta$ is measured from the slip direction leading to the positive slip sequence, which is equivalent with the twinning direction in the fcc structure.

For the hcp phase of CrCoNi, the obtained EEBs (Fig.\ref{figure1}d) indicate that both SF$^{+}_{\rm hcp}$ and SF$^{-}_{\rm hcp}$ are the primary slip modes for low ($\theta_{\rom{3}}$=0$^{\circ}$-26$^{\circ}$) and high ($\theta_{\rom{6}}$=34$^{\circ}$-60$^{\circ}$) angles, accomplished by partial dislocations with Burgers vector $ {\bf b}_{1}^{\rm hcp}$ on C  and  $ {\bf b}_{6}^{\rm hcp}$ on A  planes, respectively (Fig.\ref{figure1}f). Full dislocations (SL$^{+}_{\rm hcp}$ and SL$^{-}_{\rm hcp}$) are restricted within very narrow orientation windows, namely $\theta_{\rom{4}}$=26$^{\circ}$-30$^{\circ}$ on C and $\theta_{\rom{5}}$=30$^{\circ}$-34$^{\circ}$ on A. Hence, the hcp phase of CrCoNi is strongly prone to SF$_{\rm hcp}$  formation, i.e., to the $\varepsilon\rightarrow \gamma$ phase transformation. Notably, this phase transformation is  insensitive to crystal orientation, which is opposite to the unidirectional $\gamma\rightarrow\varepsilon$ DIMT and cTW processes in the fcc lattice \cite{McCabe2014,GUTIERREZURRUTIA20103552,LU20103079}. It is very important to recognize that the $\varepsilon\rightarrow\gamma$ transformations in the $\theta_{\rom{3}}$ and $\theta_{\rom{6}}$ regions yield fcc products exactly in the twin relationship (Fig.\ref{figure1}f), because they are realized by the positive and negative slip sequences, respectively. Therefore, such geometrical features increase the likelihood of activation of leading partials with all six Burgers vectors in the hcp crystallines during deformation, especially under severe plastic deformations due to the complex shear stresses from multiple directions \cite{Edalati2013}, which will generate nano thickness $\gamma$ and $\gamma_{\rm tw}$ lamellar simultaneously, depending on the Burgers vectors of the partials. It is worth mentioning that the presently disclosed susceptibility of hcp lattice to SF$_{\rm hcp}$ formation should play important role in the observed phase reversion from the pressure-induced hcp phase to the metastable fcc phase during decompression of CrCoNi-based high entropy alloys (HEAs)~\cite{Tracy2017,Zhang011902}.
	 
Based on the above results, we can draw a unified picture of the deformation processes for fcc CrCoNi. 
First, the pristine fcc grains with the primary slip plane having a small  $\theta$ angle (in the range of $\theta_{\rom{1}}$=0$^{\circ}$-34$^{\circ}$) deform extensively by SFs, which will be  manifested as the $\gamma\rightarrow\varepsilon$  DIMT when the maximum SF density (i.e., SFs on every other \{111\}$_{\rm fcc}$ planes) is reached with increasing strains. This step is driven by the energy decrease due to the negative $\gamma^{\rm fcc}_{\rm isf}$.  In other words, SFs avoid to form on nearest-neighboring layers to maximize the energy gain. 
Then, the deformation-induced $\varepsilon$ lamellar embedded in the fcc matrix further deform to release the local stress concentrations with increasing stains \cite{Li227,LIU2019444}. This is realized preferentially along the basal planes which serve as the primary slip planes as the \{111\}$_{\rm fcc}$ planes because of the S-N OR. Additionally, it is well documented that the basal slips are remarkably easier than other slips involving dislocations with $\langle$$\bf c$$\rangle$ character in hcp metals, particularly when $\gamma_{\rm sf}^{\rm hcp}$ is small \cite{Wu447}. Here, we measure the compatibility for activation of the leading partials of the same Burgers vector sequentially in the fcc matrix and in the hcp martensite by the relative EEB difference,  $\delta_{\rm usf}^{\rm hcp-fcc}\equiv(\gamma_{\rm usf}^{\rm hcp}-\gamma_{\rm usf}^{\rm fcc})/\gamma_{\rm usf}^{\rm fcc}$.  For CrCoNi, $\delta_{\rm usf}^{\rm hcp-fcc}$ is as small as $\sim$6\% (Table S1), which suggests that the critical resolved shear stress for the hcp basal SF formation can be easily reached at the same spot as that nucleates SF in the fcc lattice upon further loading. For $\theta$  in the range of 0$^{\circ}$-26$^{\circ}$ ($\theta_{\rom{3}}$ in Fig.\ref{figure1}f), SF$^{+}_{\rm hcp}$s are formed by the positive slips; and consequently, they transform the fresh hcp lamella $forward$ to the fcc twin, viz. ABC/ACACACAC/ABC ($\gamma/\varepsilon/\gamma$)$\rightarrow$ ABC/BA$\vert$CB$\vert$A/CAC/ABC ($\gamma/\gamma_{\rm tw}/\varepsilon/\gamma$), and thus realizing the novel TMT mechanism. With further increasing strains, full dislocations in the fcc twins are activated to accommodate deformation (Supplementary Note S1).  Immediately, the above processes explain why the strain-induced hcp phase in CrCoNi cannot grow in thickness, but remains limited to nanosize, even through it is thermodynamically stable \cite{Niu1363,Slone38,Miao35,GUO2019176}. Theoretically, if all the deformation-induced hcp lamellar with orientations in the $\theta_{\rom{3}}$ region transfer to $\gamma_{\rm tw}$, only the hcp fractions developed in the fcc grains with $\theta$ in the range of  $\theta_{\rom{1}}-\theta_{\rom{3}}$ who deform by full dislocation slips on the basal planes can survive. 
If one further considers the strong texture development and grain rotation towards twinning favorable direction during deformation~\cite{Slone38,Miao35}, very less hcp phase is expected to be reserved in actual observation. This perfectly explains why only a modest amount of hcp ($\sim$3\%) was observed at the fracture strain in CrCoNi MEA \cite{Miao35}.

\begin{figure}[h]
 \centering
 \subfloat{{\includegraphics[trim=0cm 13cm 0.1cm 3cm,clip,width=0.8\textwidth]{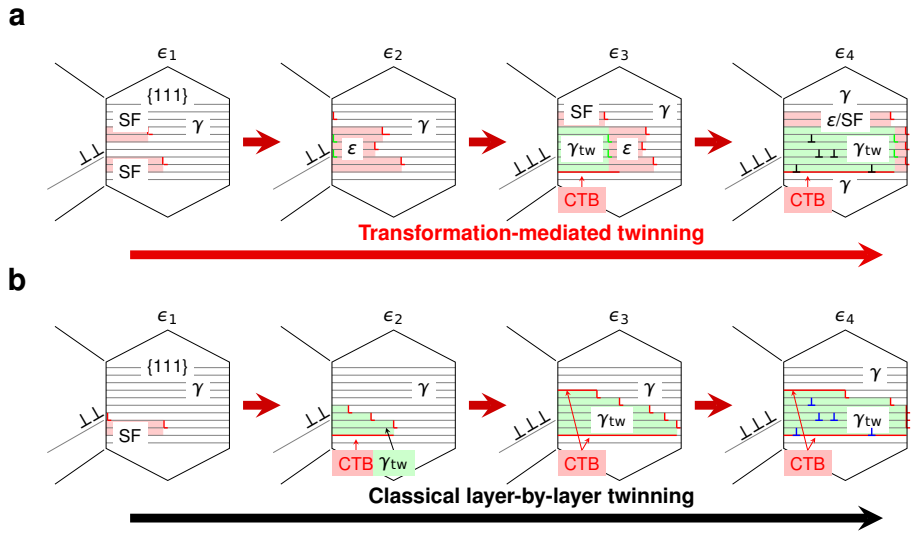}}}%
\caption {{\bf Schematics of deformation structure evolution with increasing strain levels ($\epsilon_1$ to $\epsilon_4$) via different twinning mechanisms}. {\bf a}, Transformation-mediated twinning, showing the $\gamma\rightarrow$ SF/$\varepsilon\rightarrow \gamma_{\rm tw}$ two-step twinning process, the formation of the characteristic $\gamma/\varepsilon/\gamma$ and $\gamma/\gamma_{\rm tw}/\varepsilon/\gamma$ deformation lamellar structures. {\bf b}, Classical layer-by-layer twinning in thermodynamically stable fcc materials with positive SFE.}%
\label{tmt}%
\end{figure}

The TMT mechanism provides an universal explanation for the observed $\gamma/\varepsilon/\gamma$ and $\gamma/\gamma_{\rm tw}/\varepsilon/\gamma$ lamellar structures in various metastable metals and alloys including the room-temperature fcc Co \cite{WU2005681}, Co-rich alloys \cite{REMY1976123}, stainless steels (e.g., 304) \cite{Remy1974,SHEN2012514}, high-Mn TWIP/TRIP steels \cite{REMY197799,CHOI2015391,SHEN20136093},  dual-phase fcc+hcp  Cr$_{10}$Mn$_{30}$Fe$_{50}$Co$_{10}$  \cite{Li227,Lu1804727} and single-phase CrCoNi-based  high-entropy alloys (HEAs) \cite{Lin2018236,Niu1363,Slone38,Miao35,GUO2019176}.  Noticeably, fcc twins are often observed in immediate contact with the hcp plates \cite{Niu1363,Slone38,Miao35,GUO2019176}, which cannot be incontestably solved by existing theories based on the cTW route \cite{Niu1363,Ding8919,Chen2020, Huang20182381}, but can be well expected within the TMT picture because twins are thickened through consuming newly formed SFs/hcp platelet (Fig.\ref{tmt}a). Additionally, an obvious evidence for the TMT can be found in the reported images of the deformation substructure in CrCoNi MEA obtained by high-angle annular dark-field scanning transmission electron microscopy \cite{Slone38}. Those images show that the fine hcp phase is in the immediate front of $\gamma_{\rm tw}$ in the same lamella of several atomic layers, as inferred from the present TMT mechanism (Fig.\ref{tmt}a).

When the SFE is positive, twinning nucleation and thickening prefer the cTW path in order to avoid extra energy cost (Fig.\ref{tmt}b). This is because the formation energy of coherent twin boundary (CTB) is about half of the SFE; therefore, increasing twin thickness according to the cTW mechanism does not require additional formation energy for the CTBs. In contrast, twinning according to the TMT mechanism is catalyzed by the energy gain (negative SFE) in the first step, which drives the extensive nucleation of SFs and fine $\varepsilon$ phase, both homogeneously and heterogeneously; and then, the second step of TMT, the $\varepsilon\rightarrow\gamma_{\rm tw}$ transformation, is  assisted  by the favorable basal slips in the $\varepsilon$ martensite, i.e., other slip systems in the hcp structure are much more difficult. The above facts account for the exceptional twinnability in metastable fcc alloys, which provides the general rationale for the observed relationship between the SFE and twinning prosperity relationship. Additionally, since all six slip directions on the basal planes can transfer the $\varepsilon$ martensite to $\gamma$ or $\gamma_{\rm tw}$, this characteristic of the hcp structure renders that the complicated stress environment during deformation can create dense basal plane SFs, i.e., SF$^{-}_{\rm hcp}$ and SF$^{+}_{\rm hcp}$, which divide and transfer the intermediate hcp martensite into fine scale $\gamma$ and $\gamma_{\rm tw}$ plates. In order words, the above process tends to generate dense twin and phase boundaries. Overall, these critical features of TMT enable the systems to deform sequentially by massive amount of SFs,  $\gamma\rightarrow\varepsilon$ and $\varepsilon\rightarrow\gamma/\gamma_{\rm tw}$ phase transformations, enriching the dynamic Hall-Petch effect.

\subsection{Material parameters for tailoring TMT in metastable alloys}

\begin{figure}[h]
 \centering
 \subfloat{{\includegraphics[trim=5cm 11cm 0.1cm 3cm,clip,width=0.5\textwidth]{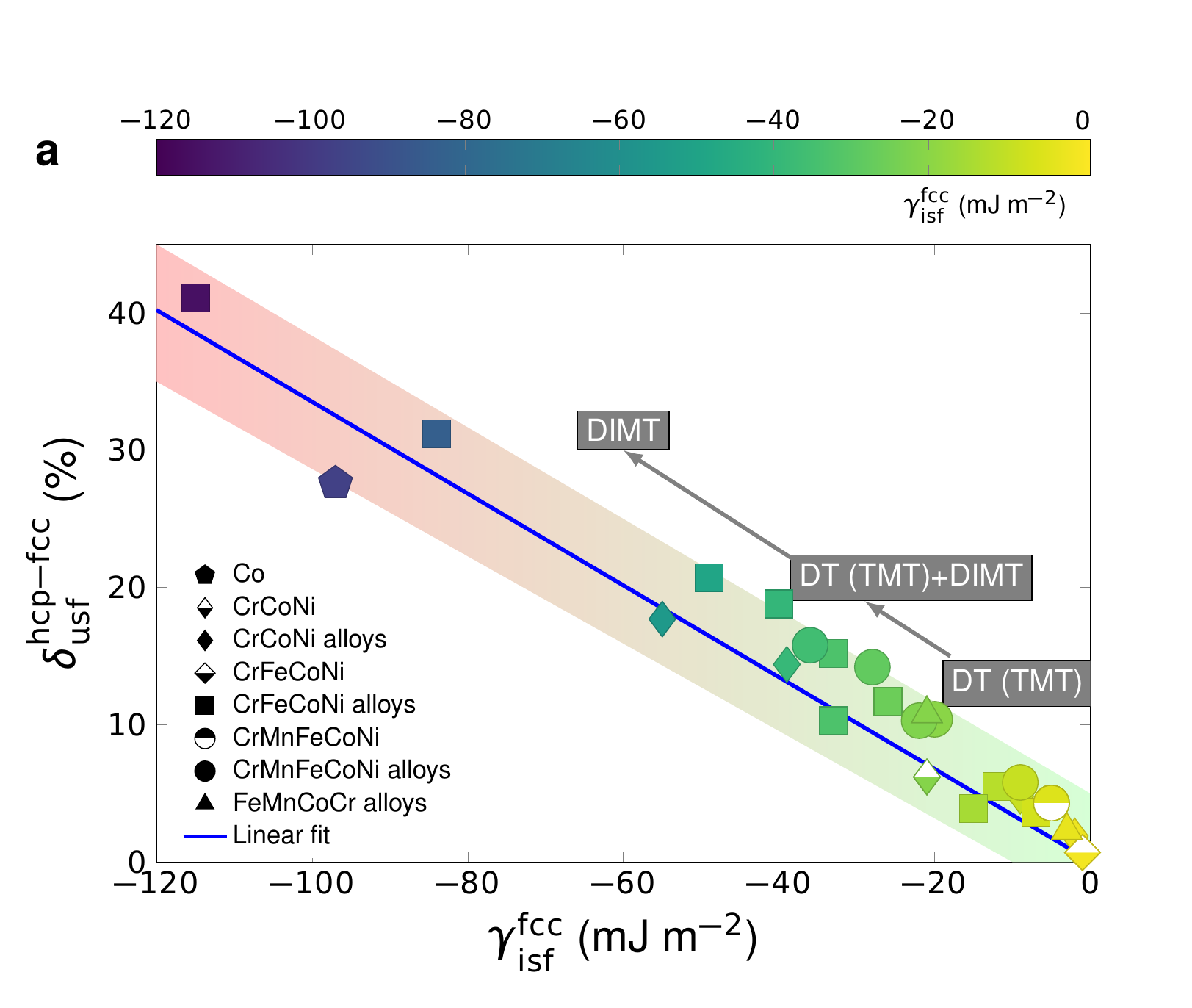}}}%
 \subfloat{{\includegraphics[trim=5cm 11cm 0.1cm 3cm,clip,width=0.5\textwidth]{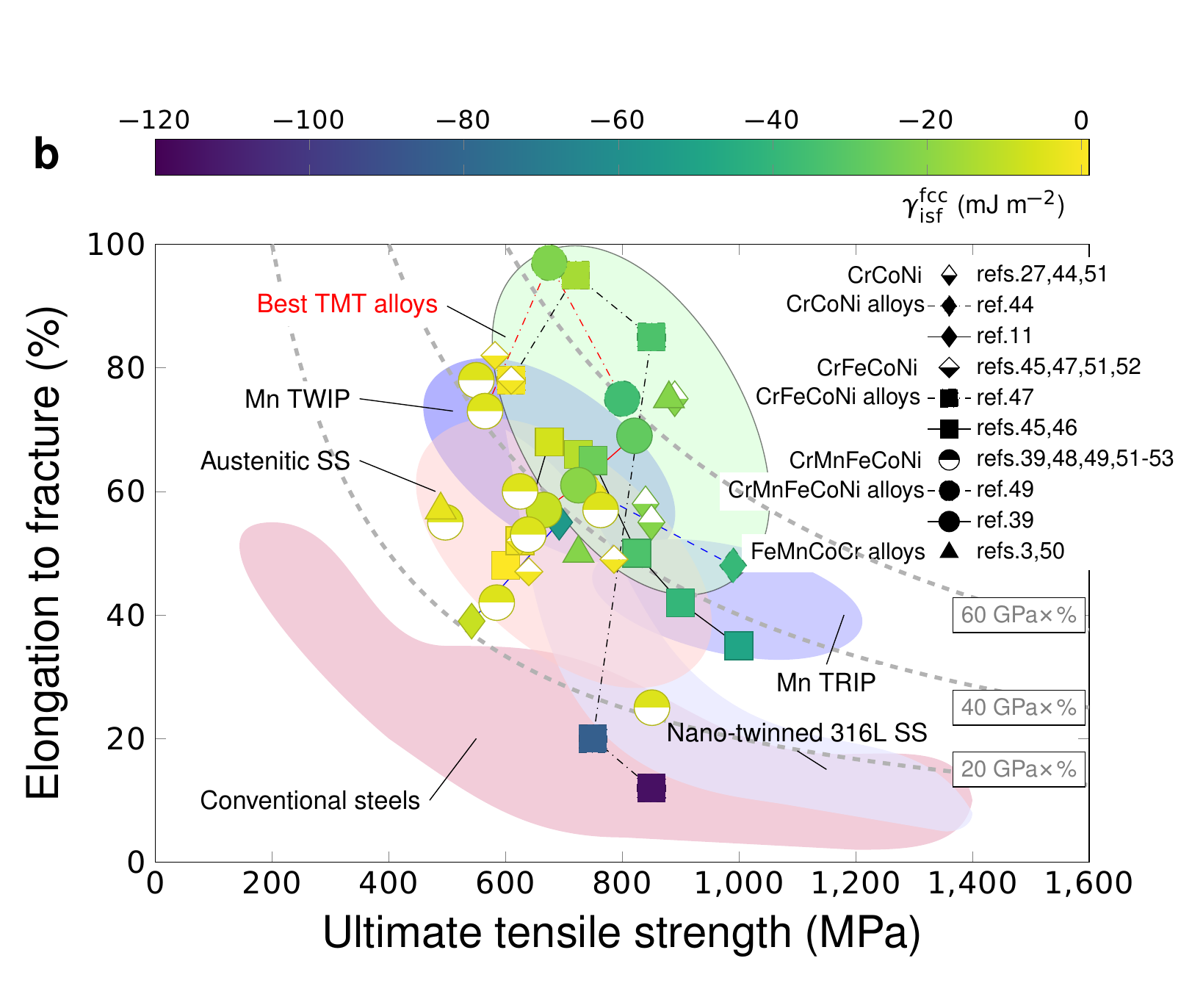}}}%
\caption{{\bf Correlations between the SFE, deformation structure and plastic performance.} {\bf a}, The calculated $\gamma^{\rm fcc}_{\rm isf}$ versus $\delta^{\rm hcp-fcc}_{\rm usf}$ relation for various metastable metals and alloys. Our TMT mechanism predicts that the primary deformation mechanism changes from DT to DIMT+DT and then to DIMT with decreasing $\gamma^{\rm fcc}_{\rm isf}$, in perfect agreement with experimental observations \cite{Slone38,Niu1363,REMY1976123,Coury2018,FANG2018221,Lin2018236,met8050369,Daixiu82,Gludovatz1153,WEI201939,Lu1804727,DENG2015124,LIU2019444}. {\bf b}, Correlation between $\gamma^{\rm fcc}_{\rm isf}$  and tensile properties. The measured room-temperature elongation to failure (EL,\%) is plotted as a function of the ultimate tensile strength (UTS, MPa) for the studied metastable alloys from Refs.\cite{Coury2018,Gludovatz2016,REMY1976123,WU2016108,ZADDACH2015373,Daixiu82,FANG2018221,met8050369,OTTO20135743,Gludovatz1153,WEI201939,LIU2019444,Li227,DENG2015124}, compared to those for conventional steels \cite{SUH201763}, austenitic stainless steels (SS) \cite{SHEN2012514,YAN20121059}, Mn TRIP and TWIP steels \cite{SUH201763,DECOOMAN2018283}. Nice correlation can be observed for the excellent mechanical properties and the SFE interval for the effectively activated TMT mechanism in ({\bf a}). The UTS$\times$EL lines of 20, 40 and 60 GPa$\times$\% are shown as thick dashed lines. All data is listed in Table S1. The color code shows the size of the corresponding $\gamma_{\rm isf}^{\rm fcc}$.}%
\label{delta}%
\end{figure}

In order to establish the general criteria for the activation of TMT, here we analyze the $\gamma^{\rm fcc}_{\rm isf}$ versus $\delta_{\rm usf}^{\rm hcp-fcc}$ relation for a group of metastable metals and alloys  (Fig.\ref{delta}a and Table S1).  It is found that $\delta_{\rm usf}^{\rm hcp-fcc}$ increases linearly with decreasing $\gamma^{\rm fcc}_{\rm isf}$. Decreasing the negative $\gamma^{\rm fcc}_{\rm isf}$ provides more thermodynamic driving force for the first step of TMT ($\gamma\rightarrow$ SF/$\varepsilon$), but simultaneously making the second step ($\varepsilon\rightarrow\gamma_{\rm tw}$) more difficult due to the higher $\gamma_{\rm sf}^{\rm hcp}$ and   $\gamma_{\rm usf}^{\rm hcp}$ which suppress the nucleation of partials at grain boundaries or the dissociation of  full dislocations on the basal planes. Therefore, the TMT mechanism is most efficient for a range of small negative $\gamma^{\rm fcc}_{\rm isf}$ accompanied by small $\delta_{\rm usf}^{\rm hcp-fcc}$. Accordingly, the primary deformation mode should gradually change from DT(TMT) to DT(TMT)+DIMT and then to DIMT with decreasing $\gamma^{\rm fcc}_{\rm isf}$, which is in perfect agreement with the experimental observation (see the prevailing deformation modes summarized in Table S1). In particular, when comparing CrCoNi with CrMnFeCoNi, we find that both have small $\delta_{\rm usf}^{\rm hcp-fcc}$ values (6.2\% and 4.3\%, respectively), the $\gamma^{\rm fcc}_{\rm isf}$ of CrCoNi (-21 mJ m$^{-2}$) is less than that of CrMnFeCoNi (-5 mJ m$^{-2}$), suggesting that the first step of TMT in CrCoNi is more pronounced than in CrMnFeCoNi (Note S2). Additionally, the larger $\delta_{\rm usf}^{\rm hcp-fcc}$ of CoCrNi than that of CrMnFeCoNi may be responsible for the larger critical twinning stress, which otherwise cannot be expected from the experimental SFEs of the two alloys \cite{LAPLANCHE2017292}. This observation clearly deciphers the underlying mechanism of the higher twinning propensity and thicker hcp nanolaths  in CrCoNi, and thus the superior mechanical performance \cite{Niu1363,LAPLANCHE2017292}. When the $\gamma^{\rm fcc}_{\rm isf}$ is further decreased, the deformation-induced $\varepsilon$ phase becomes increasingly stable and the second step of TMT will be strongly suppressed. Experimentally these alloys are usually observed as deforming by DIMT, e.g., Co, high-Co HEAs \cite{Daixiu82,LIU2019444} and stainless steels \cite{Remy1974,SHEN2012514}, but TMT still occurs in a modest fashion. In the case of metastable fcc Co with very negative SFE ($\gamma^{\rm fcc}_{\rm isf}=-97$ mJ m$^{-2}$ at 300K), $\delta_{\rm usf}^{\rm hcp-fcc}$ is as large as $\sim$28\%, which impedes the $\varepsilon\rightarrow\gamma_{\rm tw}$ transformation, and thus DT is least expected in normal tensile experiments (Note S3) \cite{WU2005681}. However, under severe deformation conditions TMT can still occur in fcc Co grains as evidenced by the characteristic $\gamma/\gamma_{\rm tw}/\varepsilon$ lamellar structure \cite{WU2005681}.

In general, high fraction of $\varepsilon$ martensite causes fast reduction of ductility by aggressively increasing the work-hardening rate which leads to premature fracture \cite{REMY197799,REMY1976123,Daixiu82}.  Therefore,  in order to reach synergy of high strength and high ductility, the present findings suggest that alloy composition should be tailored to make the best use of the TMT mechanism: decreasing the SFE the negative region to promote twinning but avoiding the formation of a large fraction of $\varepsilon$  martensite. Indeed, for all the metastable alloys investigated here (Fig.\ref{delta}b, also Table S1), the best mechanical performance is obtained for the alloys complying with this design strategy. Remarkably, the best ``TMT alloys'' possess exceptional combination of strength and ductility ($>$40-60 GPa$\times$\%), surpassing most of the conventional steels and alloys ($<$20 GPa$\times$\% \cite{SUH201763}) and even the high performance fully austenitic stainless steel (316L, $\sim$40 GPa$\times$\cite{YAN20121059}).

\section{Discussion}

\begin{figure}[h]
 \centering
 \subfloat{{\includegraphics[trim=5cm 14cm 0.1cm 3cm,clip,width=0.5\textwidth]{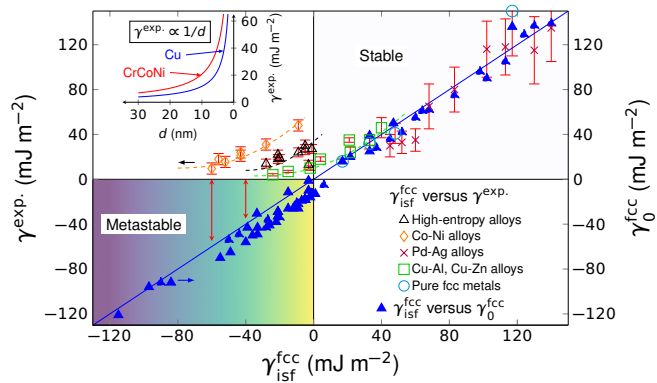}}}%
\caption {{\bf Comparison between theoretical and experimental SFEs ($\gamma_{\rm isf}^{\rm fcc}$ versus $\gamma^{\rm exp.}$)}. The relation between $\gamma^{\rm fcc}_{\rm isf}$ and $\gamma^{\rm fcc}_{0}$ for all the studied metals and alloys shows that the two quantities correlate well with each other and that decreasing $\gamma^{\rm fcc}_{\rm isf}$ indicates less stable fcc structure. The experimental SFE ($\gamma^{\rm exp.}$) is always positive and fails to reflect the true thermodynamic stability of the fcc structure in metastable alloys. The insert shows the inverse relationship between $\gamma^{\rm exp.}$ and partial separation $d$, which underlies the experimental methods for SFE measurements \cite{Reed1974,Aerts1962}. All explicit values and references are given in Tables S1 and S2.}%
\label{expsfe}%
\end{figure}

\begin{figure}[h]
 \centering
 \subfloat{{\includegraphics[width=0.5\textwidth]{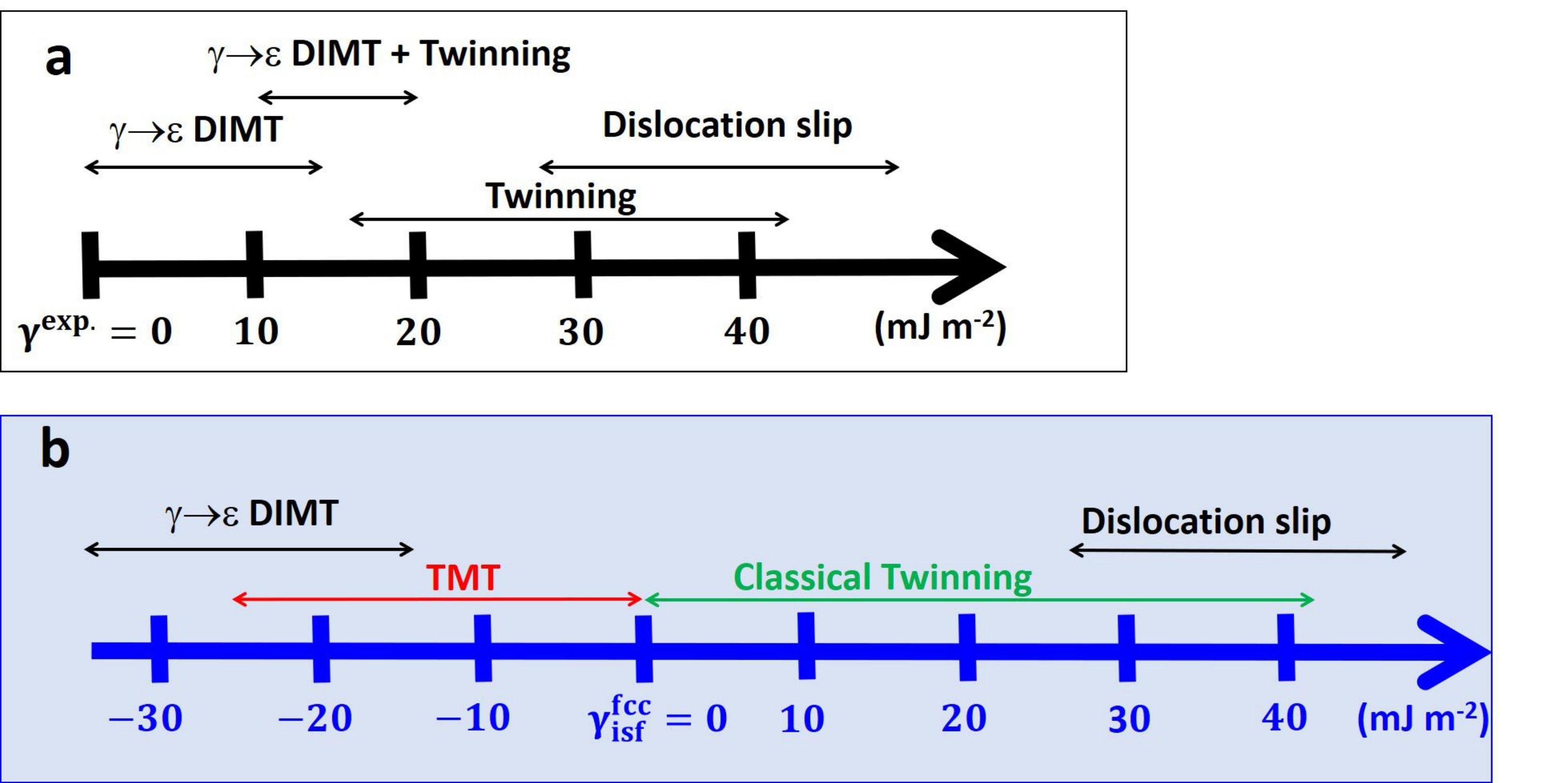}}}%
\caption {{\bf SFE - deformation mode relationship.} {\bf a}, Empirical relationship between  experimental SFE and deformation mode \cite{DECOOMAN2018283}.  {\bf b}, Revised relationship between theoretical SFE and deformation mode.}%
\label{mode}%
\end{figure}

The present findings shed light on a serious drawback met by the experimental SFE ($\gamma^{\rm exp.}$) in the realm of metastable alloys. Since an intrinsic stacking fault can be viewed as a two-layer embryo of hcp structure embedded in the fcc matrix, the SFE is traditionally connected to the hcp-fcc structural  energy difference ($\Delta G^{\gamma \rightarrow \varepsilon}$) through $\gamma^{\rm fcc}_{\rm isf}=2\Delta G^{\gamma \rightarrow \varepsilon}/A+2\sigma$, where $A$ is the SF area and $\sigma$ is the fcc/hcp coherent interfacial energy having the magnitude of  a few mJ m$^{-2}$\cite{Olson19761897}. Thus the first order approximation, $\gamma^{\rm fcc}_{0}\equiv2\Delta G^{\gamma \rightarrow \varepsilon}/A$, directly reflects the fcc vs. hcp structural stability. Indeed, an almost perfect correlation between the theoretical $\gamma_{\rm isf}^{\rm fcc}$ and $\gamma^{\rm fcc}_{0}$ can be observed in Fig.\ref{expsfe} for various stable and metastable metals and alloys. In contrast, the experimental SFEs ($\gamma^{\rm exp.}$) only agree with $\gamma_{\rm isf}^{\rm fcc}$ and $\gamma^{\rm fcc}_{0}$ in stable systems, but fail to correlate with the relative thermodynamic stability of the fcc structure in metastable materials. The trend that $\gamma^{\rm exp.}$ approaches zero with decreasing $\gamma^{\rm fcc}_{0}$ comes from its inverse relationship to the measured partial separation distance $d$ \cite{Aerts1962} or the stacking fault probability \cite{Reed1974}. Fundamentally, this failure is due to the fact that the negative excess formation energy of SF in metastable alloys can not balance the repulsive force experienced by the partials, and thus the presumed condition (i.e., positive SFE) behind all the existing experimental methods for the evaluation of the SFE breaks down \cite{Reed1974,Aerts1962}. Hence, the most critical factor affecting the nucleation and gliding of partial dislocations, i.e., the true SFE, is not accessible by the current experimental methods in metastable systems. On the other hand, DFT methods predict the SFEs consistently in both stable and metastable alloys. In addition, theory gives access to the intrinsic energy barriers for different deformation modes via the $\gamma$-surface which is also beyond the current experiments. The present findings allow us to re-establish the correlation between the SFE and the prevalent deformation modes (Fig. \ref{mode}), which paves the road to design plasticity of alloys based quantum mechanical calculations.

\section{Summary}

The disclosed TMT mechanism provides a solid physics-based and universal understanding of the exceptional twinnability and the formation of nanolamella deformation structures in metastable fcc materials. Despite that the reverse $\varepsilon\rightarrow\gamma$ transformation in hcp metals (e.g., Co \cite{Edalati2013,WU2005681}, Ti \cite{HONG2013405}, and Hf \cite{ZHAO2017271}) has long been observed and can be expected from crystallographic relationship between the two crystal structures, the two-step transformation ($\gamma\rightarrow\varepsilon\rightarrow\gamma_{\rm tw}$) has never been considered as a critical mechanism underpinning the unusual twinnability in metastable fcc alloys, particularly in TWIP steels, MEAs and HEAs. The unfortunate situation is partly because of the limitations of experiments in determining the true SFE in metastable materials, i.e.,  which has badly impeded researchers to capture the thermodynamic stability correctly. Here, using quantum-mechanical DFT calculations, we get access to the critical material parameters that controls the two-step phase transformation processes, i.e.,  $\gamma^{\rm fcc}_{\rm isf}$ and $\delta_{\rm usf}^{\rm hcp-fcc}$, with a high resolution enabling the re-establishment of the composition- SFE -deformation mechanism relationship. The present findings advance the current knowledge on the theory of plasticity in metastable fcc materials and guide the design of advanced high strength materials in the infinite composition space to overcome the strength-ductility trade-off.

\bibliographystyle{naturemag}
\bibliography{Reference}

\section*{Acknowledgments}
The authors thank Prof. Se Kyun Kwon from POSTECH for valuable discussion. H.L.Z. Thanks the National Natural Science Foundation of China (No.51871175). S.L. and L.V. acknowledge the Swedish Research Council, the Swedish Foundation for Strategic Research, the Carl Tryggers Foundations, the Swedish Foundation for International Cooperation in Research and Higher Education, the Hungarian Scientific Research Fund (OTKA 128229) for financial supports. X.S. thanks the China Scholarship Council. X.H.A. acknowledges support from Australian Research Council under DE170100053 and from The University of Sydney under the Robinson Fellowship Scheme. The computations were performed on resources provided by the Swedish National Infrastructure for Computing (SNIC) at Link\"oping. 

\section*{Author contributions}
S.L. and X.S. contributed equally to the present work. S.L., X.H.A. and L.V. initiated the study. X.S. and W.L. performed the calculations. S.L., X.S., H.L.Z. and L.V. analyzed the results and constructed the model. S.L. and L.V. wrote the manuscript and all the authors comment on the manuscript. All data is available in the main text and the supplementary materials.

\section*{Competing financial interests}
The authors declare no competing financial interests.

\section*{Supplementary materials}

Supplementary text: Note S1-S3\\

Figs. S1 to S2\\

Tables S1 and S2\\

References

\end{document}